\documentclass{article}
\usepackage{spconf,amsmath,graphicx,hyperref}
\usepackage{multirow}
\usepackage{subfig}
\usepackage{array}
\usepackage{booktabs}

\title{DDD: A Perceptually Superior Low-Response-Time DNN-based Declipper}
\name{$^*$Jayeon Yi$^1$\thanks{*This work was performed during an internship at Music and Audio Research Group, Seoul National University.}, Junghyun Koo$^2$, Kyogu Lee$^{2,3,4}$}
\address{$^1$Dept. of Electrical Engineering and Computer Science, University of Michigan \\ 
  $^2$Dept. of Intelligence and Information, Seoul National University \\
  $^3$Interdisciplinary Program in Artificial Intelligence, Seoul National University \\
  $^4$Artificial Intelligence Institute, Seoul National University }
\begin{document}
\ninept
\maketitle
\begin{abstract}
Clipping is a common nonlinear distortion that occurs whenever the input or output of an audio system exceeds the supported range. This phenomenon undermines not only the perception of speech quality but also downstream processes utilizing the disrupted signal. Therefore, a real-time-capable, robust, and low-response-time method for speech declipping (SD) is desired. In this work, we introduce DDD (Demucs-Discriminator-Declipper), a real-time-capable speech-declipping deep neural network (DNN) that requires less response time by design. We first observe that a previously untested real-time-capable DNN model, Demucs, exhibits a reasonable declipping performance. Then we utilize adversarial learning objectives to increase the perceptual quality of output speech without additional inference overhead. Subjective evaluations on harshly clipped speech shows that DDD outperforms the baselines by a wide margin in terms of speech quality. We perform detailed waveform and spectral analyses to gain an insight into the output behavior of DDD in comparison to the baselines. Finally, our streaming simulations also show that DDD is capable of sub-decisecond mean response times, outperforming the state-of-the-art DNN approach by a factor of six.

\end{abstract}
\begin{keywords}
Speech Declipping, Speech Enhancement, Adversarial Training
\end{keywords}
\section{Introduction}


Clipping occurs when an input or output of an audio system exceeds its supported range. The out-of-bounds samples are typically replaced with the maximum or minimum value, resulting in a jarring sound when played back \cite{clippingoccurswhen}. Clipping may degrade perception of human listeners or subsequent machine processing of speech \cite{clippinglisteningtest, clippingasr}. Therefore, lightweight methods to \emph{declip} speech signals are desired. 

While speech declipping (SD) has not been extensively investigated, audio declipping (AD) has been widely researched as a constrained optimization problem using non-DNN methods \cite{declippingsurvey}. Among the past approaches, the non-parametric A-SPADE \cite{spade} has shown superb reconstruction performance under low-SNR conditions ($\leq$5dB) when evaluated on music \cite{declippingsurvey}. However, when applied to speech, it was found to fall short of some DNN-based approaches \cite{upglade} in terms of speech reconstruction quality.

\begin{figure}[t]
  \centering
  \includegraphics[width=\linewidth]{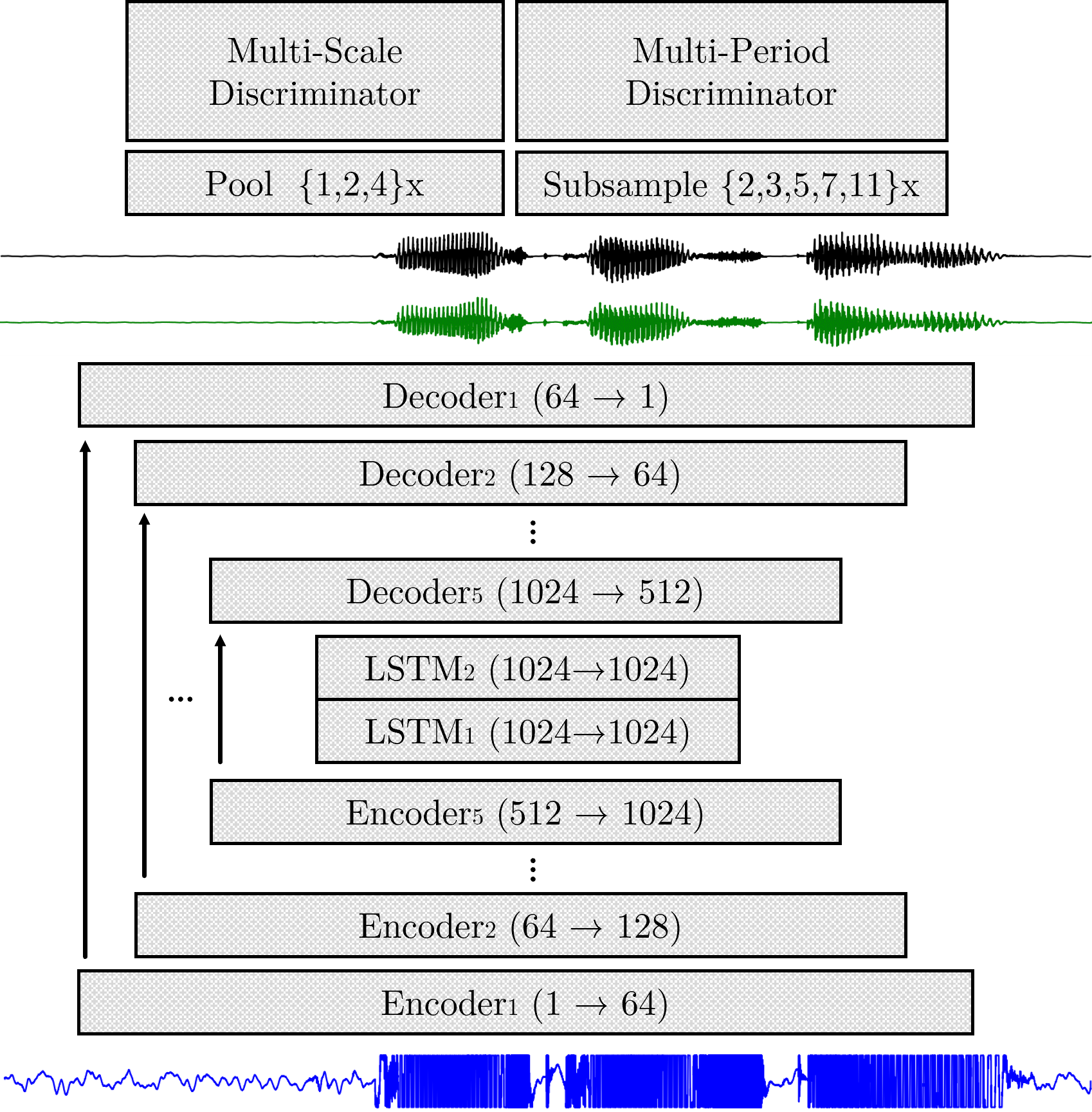}
  \caption{The architecture used to train DDD. In training, the output from our lightweight generator (green) and the original signal (black) are fed to the discriminators for an adversarial training objective and to enhance the perceptive quality of restored speech signals. The discriminators are dropped in inference and incur no overhead. } 
  \label{fig:DDD}
\end{figure}

On the other hand, although a few SD or AD methods employ deep neural networks (DNN), they often fail to outperform A-SPADE in the audio domain in terms of metrics such as $\Delta$SNR (Signal-to-Noise Ratio) \cite{applade,fourmodels}. Some other DNN-based methods reference a high number of ``future'' samples and have very high response times by design \cite{tnet}. Others require much more computation and are not real-time capable \cite{diffusion, upglade, voicefixer}. While a practitioner's first intuition may be to resort to fast speech enhancement (SE) models such as Conv-Tasnet \cite{convtasnet} or DPRNN \cite{dprnn}, our preliminary experiments showed that these models fail to converge on our SD dataset, presumably due to the fundamental differences between the additive environmental noises and the subtractive clipping noises.


To this end, we propose, train, and extensively evaluate \textbf{DDD} (\textbf{D}emucs-\textbf{D}iscriminator-\textbf{D}eclipper) - a DNN-based declipper. We train \emph{Demucs} \cite{demucs} with an adversarial training objective given by the HiFiGAN \cite{hifigan} \emph{discriminator} to better declip while preserving inference speed. Our subjective evaluations on harshly clipped speech confirm that DDD outperforms T-UNet \cite{tnet} and A-SPADE in terms of perceived audio quality. We qualitatively analyze waveforms and spectrums to investigate how the behavior of DDD differs from other DNN-based methods.
Finally, we found that our model, which is real-time capable on consumer CPUs, can be tuned to have $<$100ms response-times, one-sixth the response time compared to T-UNet.

All source code and pretrained models required to reproduce experiments are publicly available online\footnote{ \url{https://github.com/stet-stet/DDD}. Audio samples can also be heard at \url{https://stet-stet.github.io/DDD}. }.

\section{Method}


\subsection{Problem Formulation}
\label{ssec:prob_form}
Let $\mathbf{y} \in \mathbf{S}$ be an arbitrary-length speech signal, where $\mathbf{S} = \cup_{t=1}^{\infty} [-1,1]^t$. Also let us denote the length of signal $\mathbf{y}$ by $\text{len}(\mathbf{y})$; in other words, $\mathbf{y} \in [-1,1]^{\text{len}(\mathbf{y})}$. Given any threshold $\theta \in [0,1]$, we can define a \emph{hard-clip map} $f_{\theta}: \mathbf{S} \to \mathbf{S}$:
\begin{equation}
    [f_{\theta}(\mathbf{y})]_i =
    \begin{cases}
        y_i & \text{if} \hspace{.2cm} |y_i| \leq \theta \\
        \theta \cdot \text{sgn}(y_i) & \text{otherwise}
    \end{cases}
\end{equation}

Where $y_i$ is the $i$th element of $\mathbf{y}$. The aim of the SD task is to infer $\mathbf{y}$ from $f_{\theta}(\mathbf{y})$. 
This is in stark contrast to common formulations of speech enhancement (SE), where the aim is frequently to recover the original signal from its sum with an instance of roughly stochastic environmental noise.

\subsection{DDD}
Our setting for training \textbf{DDD} can be seen in Fig. \ref{fig:DDD}. We sought a twofold improvement in perceived speech quality and inference lookahead. To this end, we adopt the Generative Adversarial Network (GAN) \cite{gan} framework to introduce adversarial training objectives.

\textbf{The Generator.} We sought a model allowing for low-response-time, real-time inference on consumer-grade CPUs. We found an instance of causal Demucs \cite{demucs}, shown in Fig. \ref{fig:DDD}, to be the most suitable. The five 1D-strided-convolution blocks of the Demucs encoder each downsample the input by a factor of 4. Channels are doubled every block, with the exception of the first block which maps monochannel audio to a 64-channel representation. The decoder reverses these steps using transposed convolutions. Between the encoder and decoder is a causal 2-layer LSTM. To further reduce temporal lookahead, we upsample input and downsample output by a factor of four. Thus, the generator needs no more than 500 samples of lookahead. Although this exact setting was reported to be incapable of real-time streaming inference \cite{demucs}, we explain in section \ref{ssec:metrics} a way to accelerate inference at the expense of response time. Notably, faster SE models such as Conv-Tasnet \cite{convtasnet} or DPRNN \cite{dprnn} failed to converge in our preliminary experiments.


\textbf{The Discriminator.} We use Multi-Scale Discriminators (MSD) and Multi-Period Discriminators (MPD), which were successfully applied to speech generation \cite{hifigan} and speech enhancement \cite{hifigan2}. These discriminators are sub-stacks of strided convolution layers that take different subsets of the time-domain speech samples as input: speech subsampled with a period of \{2,3,5,7,11\} for the MPD, and \{1,2,4\}x average-pooled speech for the MSD. The discriminator is trained to label clean speech signals as 1, and restored signals as 0. We use LS-GAN \cite{lsgan} objectives for a successful training. Moreover, for the generator, we also use the feature-matching loss as outlined in \cite{melgan} and successfully applied in \cite{hifigan,hifigan2}.
 

\textbf{Training Objective.} We used the L1 loss and multi-resolution STFT loss, as follows.

\begin{equation}
    \frac{1}{T} [ \hspace{.1cm} ||\mathbf{y}-\mathbf{\hat{y}}||_1 + \sum_{i=1}^{3} L_{stft}^{(i)}(\mathbf{y},\mathbf{\hat{y}}) \hspace{.1cm} ]
\end{equation}
where $\mathbf{y}$ is the ground truth, $\hat{\mathbf{y}}$ is the network output, and



\begin{align}
    L_{stft}^{(i)}(\mathbf{y},\mathbf{\hat{y}}) &= \frac{|| \hspace{.1cm} |\text{STFT}^{(i)}(\mathbf{y})| - |\text{STFT}^{(i)}(\mathbf{\hat{y}})| \hspace{.1cm} ||_F}{|| \hspace{.1cm} |\text{STFT}^{(i)}(\mathbf{y})| \hspace{.1cm} ||_F} \nonumber \\
                                                &+ ||\log |\text{STFT}^{(i)}(\mathbf{y})| - \log |\text{STFT}^{(i)}(\mathbf{\hat{y}})| \hspace{.1cm} ||_1 ,
\end{align}
where $|| \cdot ||_1$ and $|| \cdot ||_F$ each refers to the L1 and Frobenius norm. FFT sizes of 512, 1024, and 2048 were used for each $\text{STFT}^{(i)}$. The final training objective for the generator was a sum of the above, the feature matching loss multiplied by four, and the discriminator loss.


\begin{figure*}[t] 
  \centering
  \includegraphics[width=\textwidth]{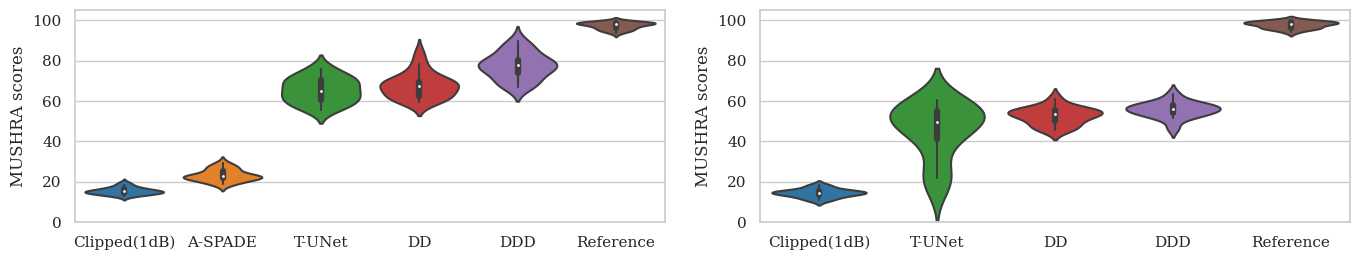} \\
  \caption{Violin plots of subjective evaluation results with VBDM-1dB-Testset (left) and DNS-1dB-Testset (right). }
  \label{fig:subjective_results}
\end{figure*}

\begin{figure*}[t]
  \centering
  \captionsetup[subfigure]{justification=centering}
  \subfloat[\label{subfig:qualitative_waveform}Typical waveform of clean and declipped speech. The two gray horizontal lines denote the clipping thresholds 
   used. ]{\includegraphics[width=\textwidth]{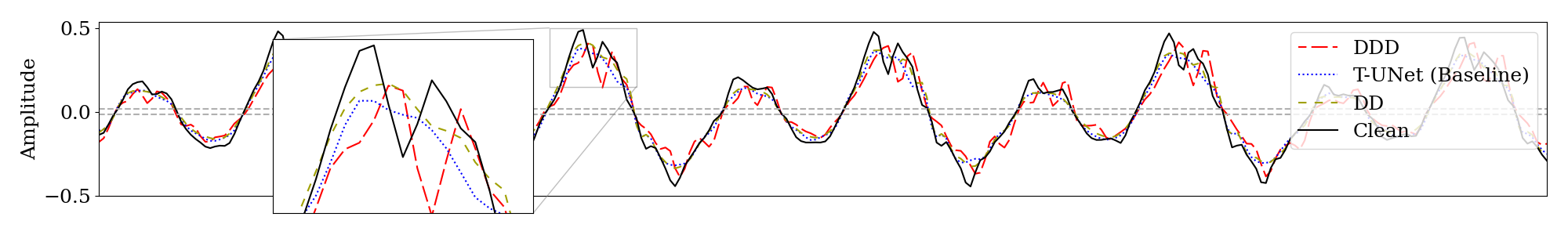} } 
  \\
  \subfloat[\label{subfig:qualitative_spectrum}Spectrum of the region in (a) without LUFS-normalization. DD and T-UNet often exhibits a flat, low spectrum beyond 3-4kHz, \\ 
  failing to model any formants in the area. ]{\includegraphics[width=\textwidth]{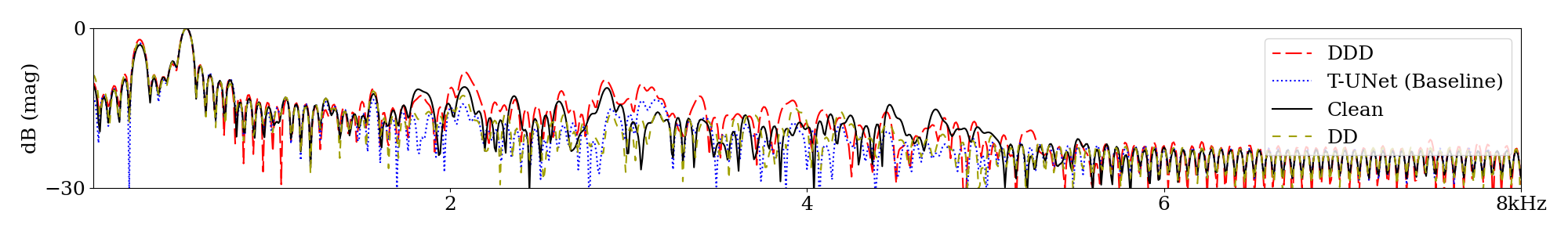} }
  
  \caption{ Reconstructed results from a hard-clipped signal (SNR=1dB). DDD, T-UNet, and DD reconstruction results are shown with the original clean speech. DDD-declipped speech shows elements of natural speech that baseline T-UNet or DD do not exhibit. The DD/baseline-declipped waveforms typically fail to recreate (a) the ``spiky'' contours of clean speech as well as (b) some higher-order formants.}
  \label{fig:qualitative}
\end{figure*}

\section{Experimental Setup}

\subsection{Dataset \& Preprocessing}

The ``clean'' splits of the Voicebank-DEMAND dataset \cite{vbdm}, comprised of a ``train" split of length 9.4 hours and a ``test" split of 35 mins, were downsampled to 16kHz. In training, the ``train'' split samples were clipped on-the-fly with a random threshold and aligned with the unprocessed speech. To place more emphasis on heavily clipped speech, we sampled $s$ uniformly from $[-2.0,-0.9]$ and used $\theta=10^s$ as the final threshold. The range of $\theta$ covered SNRs of approximately 1dB to 9dB.

The 150 clean ``test''-split utterances of Interspeech 2020 DNS-Challenge dataset \cite{dns2020interspeech}, also downsampled to 16kHz, were used together with the ``test'' split of Voicebank-DEMAND for evaluation. We clip these sets, henceforth ``VBDM-Testset'' and ``DNS-Testset'', to have SNRs of 1dB, 3dB, 7dB, and 15dB, calculating the SNR as

\begin{equation}
    \text{SNR} = 10 \log_{10}{\frac{||\mathbf{y}||^2}{||\mathbf{x}-\mathbf{y}||^2}},
\end{equation}
where $\mathbf{y}$ and $\mathbf{x}$ represent the clean and noisy signals. We denote each of these sets as ``\{VBDM,DNS\}-\textit{X}dB-Testset'' where \textit{X} is the SNR.

\subsection{Baseline Approaches}

As a DNN baseline, we re-implemented and trained T-UNet \cite{tnet}, the state-of-the-art for the SD task. This architecture employs strided convolutions to downsample input speech by a factor of two every layer, then upsamples with sub-pixel convolutions to generate raw waveforms. We do not use the discriminator for T-UNet. As a non-DNN baseline, we use A-SPADE which has shown competitive performance in AD \cite{fourmodels}. We use a MATLAB implementation provided by a past work \cite{declippingsurvey}. Finally, as an ablation, we train the generator without the discriminator. We refer to the trained model as \textbf{DD}. 

\subsection{Network Training Details}

For the T-UNet, we followed the training and inference procedures given in the original paper \cite{tnet}. We padded and split all utterances into $2^{14}$-sample-segments to make them compatible with the T-UNet. In inference, the segments were stitched back together. For DD and DDD, 24,000-samples-long utterances were used. Inference was performed without any segmentation.

All models are trained for a total of 75 epochs with the AdamW optimizer with a learning rate of $10^{-4}$, $\beta_1 = 0.9$, $\beta_2 = 0.999$, and a weight decay of $10^{-2}$. A batch size of 32 was used for DD and T-UNet, and 2 was used for DDD. 


\begin{table*}[t]
\centering
\caption{All models were trained on the VoiceBank-DEMAND dataset and evaluated on datasets indicated. The proposed models have faster response times than baseline approaches. PESQ and STOI(\%) are denoted for each set and input SNR. Notably, DDD was trained in a GAN framework, so the objective metrics below may not fully represent the perceptual audio quality - which may be better illustrated by Fig. \ref{fig:subjective_results}. }
\begin{tabular}{c c c c c c c c c c} 
 \toprule
   & \multicolumn{4}{c}{SNR (VoiceBank-DEMAND) (dB)} & \multicolumn{3}{c}{SNR (DNS-Challenge) (dB)} & \multicolumn{2}{c}{Streaming Simulations} \\ 
 \cmidrule(lr){2-5}\cmidrule(lr){6-8}\cmidrule(lr){9-10}
  Method & 1 & 3 & 7 & 15 & 1 & 3 & 7 & RTF & Response (ms) \\ [0.5ex] 
  \midrule
  \midrule
 Clipped & 1.15 / 78 & 1.39 / 86 & 2.02 / 94 & 3.20 / 98 & 1.14 / 70 & 1.33 / 82 & 1.93 / 92 & - & -\\
  \midrule
 A-SPADE \cite{spade} & 1.54 / 80 & 2.02 / 90 & 2.85 / 96 & 3.89 / 99 & 1.37 / 76 & 1.73 / 87 & 2.66 / 96 & $>$10 & -\\ 
 T-UNet \cite{tnet} & 2.97 / 94 & \textbf{3.40} / 97 & 3.84 / 98 & 4.18 / 99 & 2.13 / 87 & \textbf{2.87} / 93 & \textbf{3.61} / 97 & \textbf{0.634} & 544 \\
  \midrule 
 DD (Ablation) & \textbf{2.99} / 94 & \textbf{3.40} / 97 & \textbf{3.85} / 98 & \textbf{4.24} / 100 & \textbf{2.16} / 88 & 2.78 / 93 & 3.54 / 97 & 0.881 & \textbf{88} \\
 DDD (Proposed) & 2.55 / 93 & 3.27 / 96 & \textbf{3.85} / 98 & 4.20 / 99 & 1.82 / 86 & 2.48 / 92 & 3.36 / 96 & 0.881 & \textbf{88} \\
 \bottomrule
 \label{tab:objective_results}
\end{tabular}
\end{table*}

\subsection{Evaluation Metrics}
\label{ssec:metrics}

\noindent \textbf{Subjective Evaluation.} Subjective tests were performed to more accurately gauge absolute speech quality.
We used the webMUSHRA framework \cite{webmushra} to run a MUSHRA-like test \cite{mushra}. 20 and 15 samples randomly selected from the VDBM-1dB-Testset and DNS-1dB-Testset were run through the trained models. Afterwards, the inputs, outputs, and the associated clean samples were normalized to -27 LUFS. 24 participants with more than two years of experience in music production or audio engineering were asked to score each normalized sample on a 100-point scale based on the quality of speech. 
Alteration of speech content, if any, was asked to be considered. Most participants were unfamiliar with hearing tests. We filtered out respondents who failed to grasp the existence of a hidden reference within the first four sets via post-survey interviews.

We excluded A-SPADE outputs from the longer 15 DNS-Testset questions to minimize ear fatigue. This exclusion in only one dataset can be justified since A-SPADE is a non-parametric model.

\noindent \textbf{Inference Time Evaluation.} We ran inferences on the VDBM-Testset using a batch size of 1, and divided the total inference times for each approach by the total set duration to calculate the throughput. We also calculated the MACs (Multiply-ACcumulate operations) per sample for each DNN-based approach. 

To measure response times and the real-time factors (RTF), we simulated a typical audio streaming application. We fed samples into the model at a rate of 16,000 samples/sec for 100 seconds, then calculated the mean time interval between the feeding and output of every 500$^\text{th}$ sample. Samples were buffered, and then fed into the network once a sufficient number of samples were gathered to yield output. Due to its architecture, the T-UNet needs at least $2^{14}$ samples to generate an output - therefore $2^{14}$ samples were buffered before yielding any output. DD and DDD needs much less, only requiring enough to make one input vector to feed the LSTM. To enable out model to run real-time in streaming settings, we buffered enough samples to produce four input vectors for the LSTM. In this case, the required lookahead is 1,429 samples, due to how we upsample inputs by a factor of four. This evaluation was performed on a Ryzen 7 5700X CPU @ 3.4GHz. 

\noindent \textbf{Objective Evaluation.} Two objective metrics were employed to evaluate speech: PESQ (Perceptual Evaluation of Speech Quality) \cite{pesq} and STOI (Short-Time Objective Intelligibility) \cite{stoi}.  PESQ is a similarity-based metric that predicts speech quality, and is given as a score between -0.5 and 4.5. STOI is likewise a similarity-based objective that aims to predict the intelligibility of speech. Although they were recently singled out to be unreliable as a measure of absolute speech quality due to how they take speech similarity into account \cite{unreliable}, both metrics have seen popular use in SD literature \cite{declippingdeepflitering, tnet}. Following common practice, we report these metrics for a variety of input SNR values.


\section{Results}

\subsection{Subjective Evaluations}
\label{ssec:subjective}
Subjective evaluation results are presented in Fig. \ref{fig:subjective_results}.  
Out of twenty-four responses, four were filtered out as described in Section \ref{ssec:metrics}. 
To account for each participant's different scoring criteria, the 20 scores for each samples were averaged before drawing the violin plot.
A pairwise post-hoc paired t-test with Bonferroni correction concludes that differences between DDD and all other models for both datasets are statistically significant ($p < 0.0001$). While still distinguishable from the hidden reference, DDD may be a significant improvement over the T-UNet or DD. Moreover, comparing DD and DDD suggests that the adversarial training objectives may have helped boost the quality of speech produced by DDD.



The scores for the DNS-Testset are smaller than those for the Voicebank-DEMAND dataset - a natural result, as the models were trained on the latter. Thus, the DNS-Challenge evaluations were in effect zero-shot evaluations on utterances recorded in a different environment. The long "tail" of score distributions exhibited by T-UNet on this set hints at some catastrophic failures of the T-UNet on unseen recording environments. This result implies that DDD is more robust compared to the T-UNet.

\subsection{Qualitative Analysis}
\label{ssec:qualitative}

In Fig. \ref{subfig:qualitative_waveform}, a typical waveform of clean and clipped (1dB) speech is presented with speech declipped by T-UNet, DD, and DDD. Notably, T-UNet and DD are often unable to reconstruct the up-and-down ``ruggedness" of the clean-speech waveform as shown in Fig. \ref{subfig:qualitative_waveform}.

In the regions that were previously saturated, the DD and T-UNet outputs are less ``rugged" than the clean signal; indeed, over the entire VBDM-1dB-Testset, the associated clean signals have a total of 1.52M local extrema over the previously saturated regions, about 35-50\% more than DD (1.17M extrema) or T-UNet outputs (1.09M extrema). This problem is alleviated by DDD (1.67M extrema). Therefore, we suspect that the inability of DD and T-UNet to achieve a perfect reconstruction of speech is at least partially associated with their incapacity to reproduce the up-and-down ``ruggedness" of the clean-speech waveform.

Fig. \ref{subfig:qualitative_spectrum} shows the Fourier transform coefficients of the same region. Across the entire dataset, all model outputs conform to the original spectrum in the sub-1kHz regions. However, the T-UNet frequently fails to reproduce higher-order formants at 3kHz and beyond. In contrast, DDD does seem to model higher-frequency(2-4kHz) regions, sometimes resulting in formants minutely different from the clean speech but resulting in coherent and high-quality speech nonetheless. These observations may explain the higher subjective scores of DDD as presented in Fig. \ref{fig:subjective_results}.

\subsection{Objective Evaluations}
\label{ssec:objective}

Table \ref{tab:objective_results} presents objective evaluation results. DD performs comparably to T-UNet in terms of similarity-based metrics, with DDD following close behind, and A-SPADE scoring far lower. However, we cannot conclude that T-UNet or DD provides better reconstructions - subjective evaluations are known to be more reliable \cite{mushra}. Moreover, similarity-based scores have been previously pointed out to be more unreliable for adversarially-trained speech networks such as DDD \cite{hifigan2, unreliable}. Reflecting on previous analyses, we may conclude that DDD has traded away an acceptable degree of dataset conformance for perceptual quality. 

RTF and response time measurement results are presented in the rightmost columns of Table \ref{tab:objective_results}. MACs per sample and the throughput with respect to temporal length were measured as well: 0.19M/sample and 0.15x for T-UNet, and 0.48M/sample and 0.34x for DD or DDD. As noted in past works \cite{declippingsurvey}, the RTF for A-SPADE differs with input SNR, but was typically over 20. While both DNN-based approaches have a sub-unity RTF, T-UNet has a mean response time of 0.58 seconds, which is unacceptably high for many applications. In contrast, DD and DDD have sub-decisecond mean response times. Moreover, as discussed in section \ref{ssec:metrics} DD and DDD can trade RTF and response time using buffers: it can, for instance, be configured to match the RTF of T-UNet and still yield one third the response time. Therefore, DD and DDD may be better suited to real-time audio processing.






\section{Conclusion}

In this work we proposed and trained DDD, a DNN model capable of real-time, low-response-time, high-quality declipping. We adopted the GAN framework along with many tricks to boost the capabilities of Demucs \cite{demucs}. MUSHRA-like subjective evaluations on harshly clipped speech revealed that DDD outperforms previously proposed methods by a large margin. We performed qualitative analyses, observing how existing approaches suffer from ``round-waveform" behavior accompanied by a neglection of high-frequency modeling. Finally, DDD was found to exhibit a response time far lower than the previous state-of-the-art. While advarsarial training objectives enabled a perceptually-better speech declipping with lower response times, an exact declipping of speech remains an open problem.

\bibliographystyle{IEEEbib}
\bibliography{strings}

\end{document}